\documentclass[11pt]{article}
\usepackage[margin=1in]{geometry}
\usepackage{graphicx}

\def\beq{\begin{equation}}
\def\eeq#1{\label{#1}\end{equation}}
\def\eeqn{\end{equation}}

\def\beqa{\begin{eqnarray}}
\def\eeqa#1{\label{#1}\end{eqnarray}}
\def\eeqan{\end{eqnarray}}

\let\bar=\overbar

\def\Dslash{\not{\hbox{\kern-4pt $D$}}}
\def\dslash{\not{\hbox{\kern-2pt $\del$}}}

\def\msb{{\bar{\ssstyle M \kern -1pt S}}}

\def\Title#1{\begin{center} {\Large {\bf #1} } \end{center}}
\def\Author#1{\begin{center} {\normalsize {\sc #1} } \end{center}}
\def\Institution#1{\begin{center} {\normalsize {\it #1} } \end{center}}
\def\Abstract#1{\noindent {\normalsize {\bf Abstract:} {\normalfont #1}}}
\def\Conference{\vspace{4mm}\begin{raggedright} {\normalsize {\it Talk presented at the 2019 Meeting of the Division of Particles and Fields of the American Physical Society (DPF2019), July 29--August 2, 2019, Northeastern University, Boston, C1907293.} } \end{raggedright}\vspace{4mm}}

\begin{document}

%
%

\Title{Search for Standard Model Production of Four Top Quarks}

\Author{Caleb Fangmeier}

\Institution{Department of Physics and Astronomy\\ University of Nebraska-Lincoln}

\Abstract{This talk describes efforts towards a first measurement of the standard model production of four top quarks with results based on up to the full LHC Run 2 dataset collected at CMS at $\sqrt{s}=13\mathrm{TeV}$. It includes interpretations of this measurement to constrain properties of the Higgs Boson and new physics scenarios including dark matter.}

\Conference%

%
%

\section{Introduction}

\begin{figure}[htb]
\centering
\includegraphics[height=1.5in]{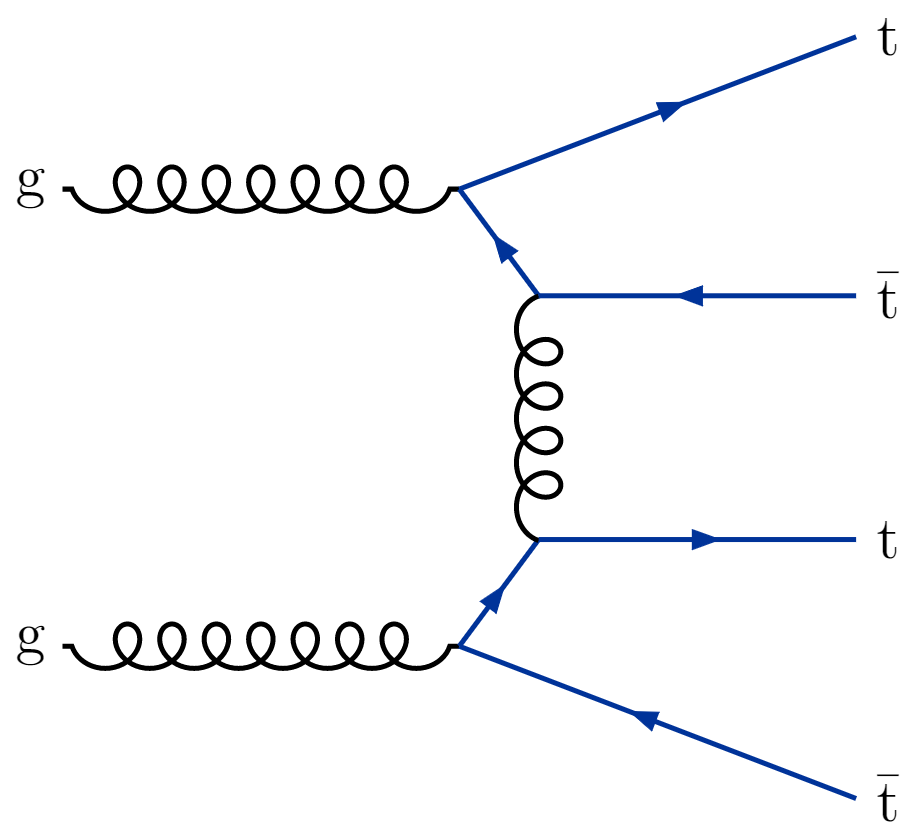}
\includegraphics[height=1.5in]{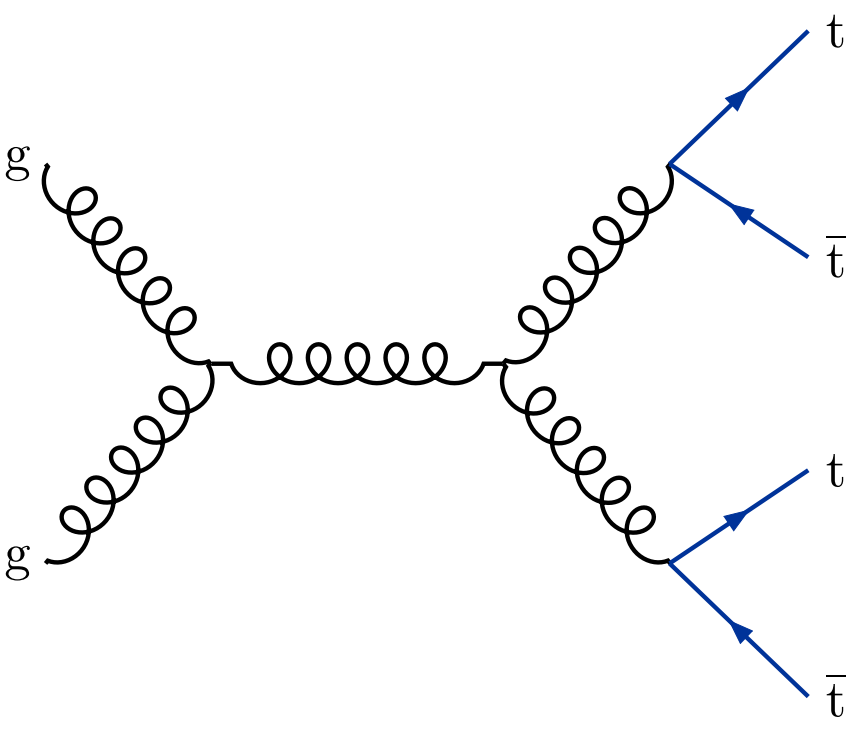}
\caption{Some representative leading order QCD production diagrams for four top quarks}\label{fig:feyn_qcd}
\end{figure}

Four top quark production is a rare Standard Model process with a cross section of $\sigma\left(\mathrm{pp}\rightarrow t\bar{t}t\bar{t}\right)=12~\mathrm{fb}$ calculated at next-to-leading order (NLO) at 13~TeV center of mass energy~\cite{Frederix:2017wme}. The dominant production mode of $t\bar{t}t\bar{t}$ at the LHC is through QCD diagrams such as those shown in figure~\ref{fig:feyn_qcd}. There are also smaller contributions from Higgs and vector boson mediated diagrams.

The $t\bar{t}t\bar{t}$ cross section can be used to constrain Standard Model parameters such as the top quark Yukawa coupling, as well as properties of several Beyond the Standard Model theories; most notably the Type-II two-Higgs doublet models (2HDM), simplified dark matter, and off-shell mediators such as a top phillic $Z'$.

Four top quark events are notable for their large number of potential final states. Depending on how the W bosons in the event decay, there can be from zero to four prompt leptons and from twelve to four jets, of which four will originate from b quark decays. Covering all final states would be too complicated for a single analysis, so events are classified by the multiplicity and relative charge of final state leptons, with different categories having dedicated analyses. This analysis covers the case where there is a same-sign pair of leptons, including three or more leptons. This report will give a summary of the analysis, but additional details can be found in~\cite{1908.06463}.

\section{Event Selection and Background Estimation}

\begin{figure}[htb]
  \centering
  \includegraphics[height=2.0in]{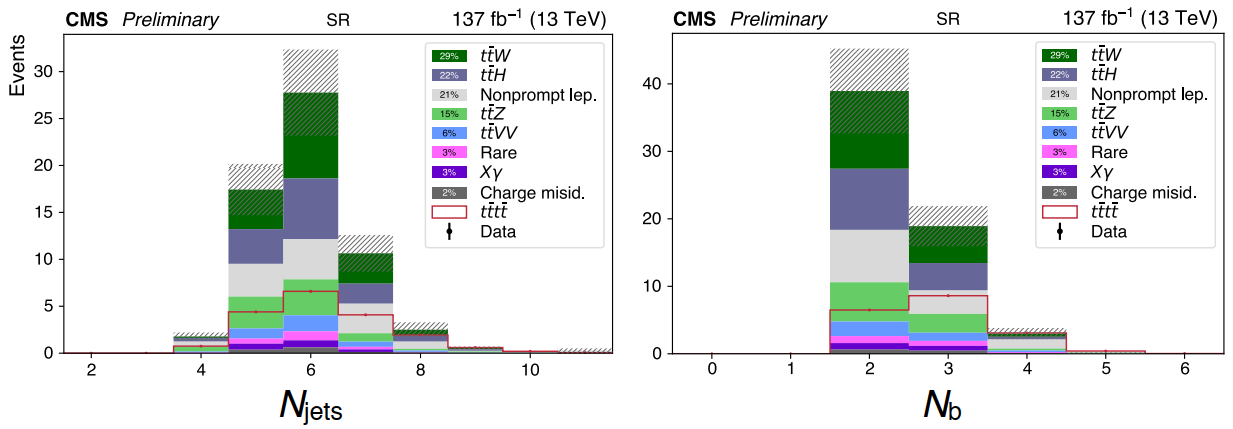}
  \caption{Content of events within the baseline selection\cite{1908.06463}}\label{fig:baseline}
\end{figure}

The baseline selection is designed to efficiently select $t\bar{t}t\bar{t}$ events with as small as possible contribution of background processes. We first require a high quality (or ``tight'') same-sign lepton pair, or three or more tight leptons. Lepton here means electrons, muons, and leptonically decaying $\tau$. The highest $p_T$ lepton is required to have at least 25~GeV of transverse momentum while all others have a more relaxed requirement of 20~GeV. The event must also contain at least two jets with $p_T$ greater than 40~GeV, and at least two jets identified as originating from a bottom quark with $p_T$ over 25~GeV. Furthermore, we require the scalar sum of transverse jet energy to be greater than 300~GeV, and at least 50~GeV in missing transverse momentum. Finally, if the event contains an additional lepton satisfying a relaxed ID requirement that forms an opposite-sign same-flavor pair with one of the tight leptons, and the pair has an invariant mass within 15~GeV of $m_Z$ then the event is discarded. If there is an additional tight lepton that fulfills these requirements, then the event is instead placed in a dedicated $t\bar{t}Z$ control region called CRZ\@. Figure~\ref{fig:baseline} shows the expected contribution of different processes to the baseline selection differentially in the number of jets and number of b-tagged jets.

The backgrounds for this analysis come in two types: processes with genuine prompt same-sign lepton pairs, and events with ``fake'' same-sign lepton pairs. Of the former, the most important processes are $t\bar{t}W$, $t\bar{t}H$, and $t\bar{t}Z$. The contributions from these processes are estimated by simulating events at NLO\@. In the case of $t\bar{t}Z$, there is a dedicated control region that is used to normalize its contribution.

The contribution of the nonprompt lepton background is estimated using the ``tight-to-loose'' ratio method~\cite{SUS-15-008}. This is a common technique employed in many analyses which deal with backgrounds resulting from nonprompt leptons. The method first measures the proportion of lower quality or ``loose'' leptons that also pass the stricter tight requirements in a sideband that is enriched in nonprompt leptons. This proportion can then be used to calculate a weight that is applied to events that would pass the baseline selection except for having one tight and one loose-not-tight lepton instead of the two tight leptons normally required. These weighted events make up the nonprompt background estimation. The charge misidentified background estimate works similarly except that the charge misidentification probability is measured in simulation, and the resulting transfer factor is applied to opposite-sign events.

\section{Results}

\begin{figure}[htb]
\centering
\includegraphics[height=2.5in]{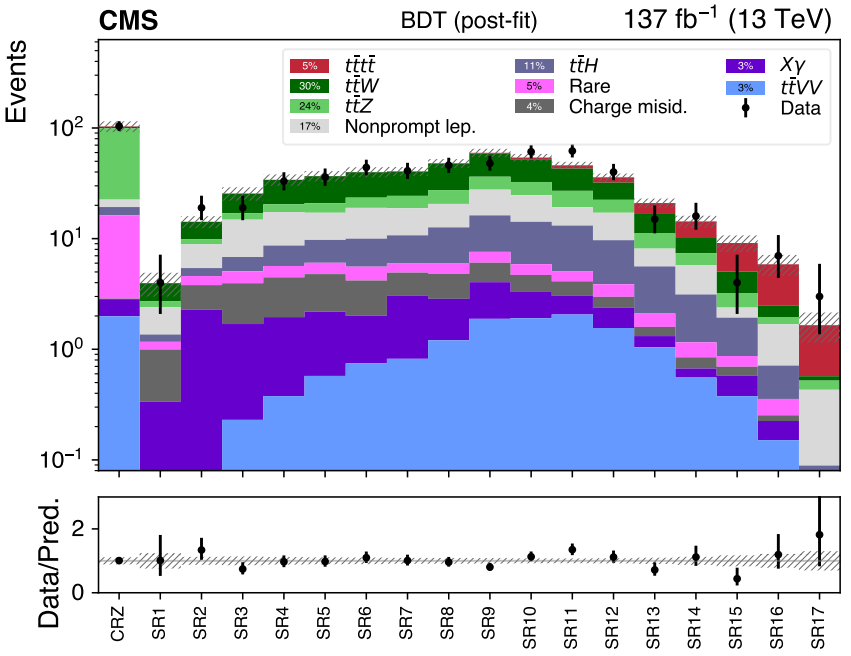}
\caption{Observed yields of the binned BDT discriminant compared to the post-fit predictions for signal and background processes. The hatched area shows the total uncertainty on the background and signal prediction.\cite{1908.06463}}
\label{fig:bdt}
\end{figure}

A boosted decision tree (BDT) classifier is utilized to separate $t\bar{t}t\bar{t}$ events from background events. The BDT consists of 500 trees with a depth of 4. It uses 19 event level variables which take into account object multiplicities, reconstruction quality, energies, and angular relations. The results of applying this BDT to events in the signal region are shown in figure~\ref{fig:bdt}.

A binned likelihood is constructed from 17 bins of the BDT discriminant with an additional bin containing CRZ\@. A profile maximum-likelihood fit is then performed with experimental and theoretical uncertainties incorporated as nuisance parameters. The fit results in a measured $t\bar{t}t\bar{t}$ cross section 
of $12.6^{+5.8}_{-5.8}$~fb with a 68\% confidence interval. A cross check was performed using an event binning based on the number of jets, b jets, and leptons instead of the BDT discriminant. This yielded a cross section of $9.4^{+6.2}_{-5.6}$~fb, consistent with the BDT-based measurement.

\section{Interpretations}

\begin{figure}[htb]
\centering
\includegraphics[height=2.5in]{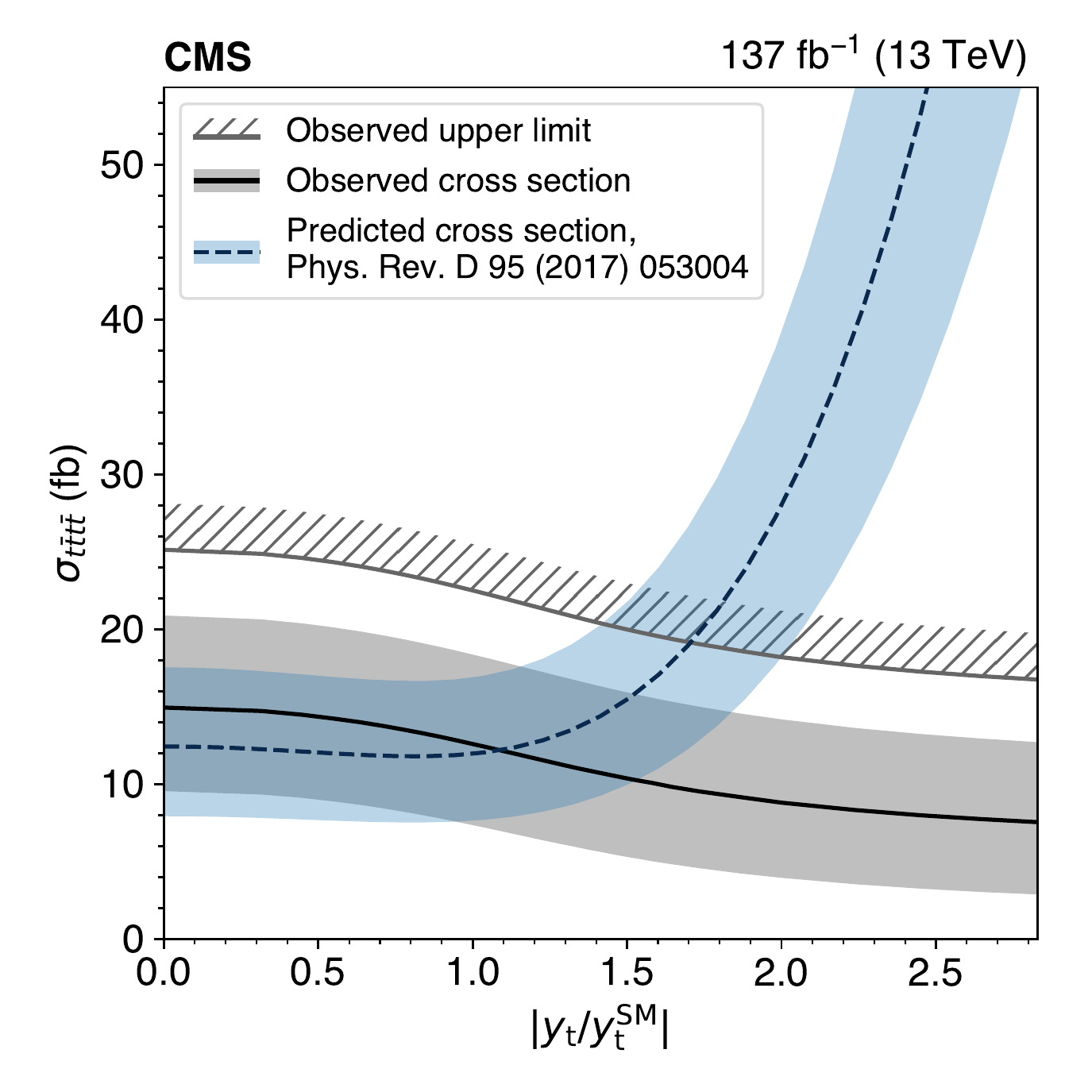}
\caption{The observed $\sigma\left(t\bar{t}t\bar{t}\right)$ and 95\% CL upper limit as a function of $|y_t/y_t^{\mathrm{SM}}|$. The dashed line shows the predicted cross section calculated at LO and scaled to the NLO result of Ref.~\cite{Frederix:2017wme}. The observed limit on $t\bar{t}t\bar{t}$ varies as a function of $|y_t/y_t^{\mathrm{SM}}|$ because the $t\bar{t}H$ ($H\rightarrow WW$) background grows with increasing $y_t$.\cite{1908.06463}}
\label{fig:yukawa}
\end{figure}

The result of the analysis can be used to constrain Standard Model parameters, as well as BSM processes that can affect the $t\bar{t}t\bar{t}$ production rate. The existence of off-shell Higgs mediated Feynman diagrams for $t\bar{t}t\bar{t}$ production means that the cross section is sensitive to the top quark Yukawa coupling~\cite{TopYukawaTTTT, TopYukawaTTTTnew}. Figure~\ref{fig:yukawa} shows the predicted cross section as a function of the ratio of the top Yukawa coupling to its Standard Model value. We observe a limit of $|y_t/y_t^{\mathrm{SM}}|<1.7$ at the 95\% confidence level.

\begin{figure}[h!]
\centering
\includegraphics[height=2.5in]{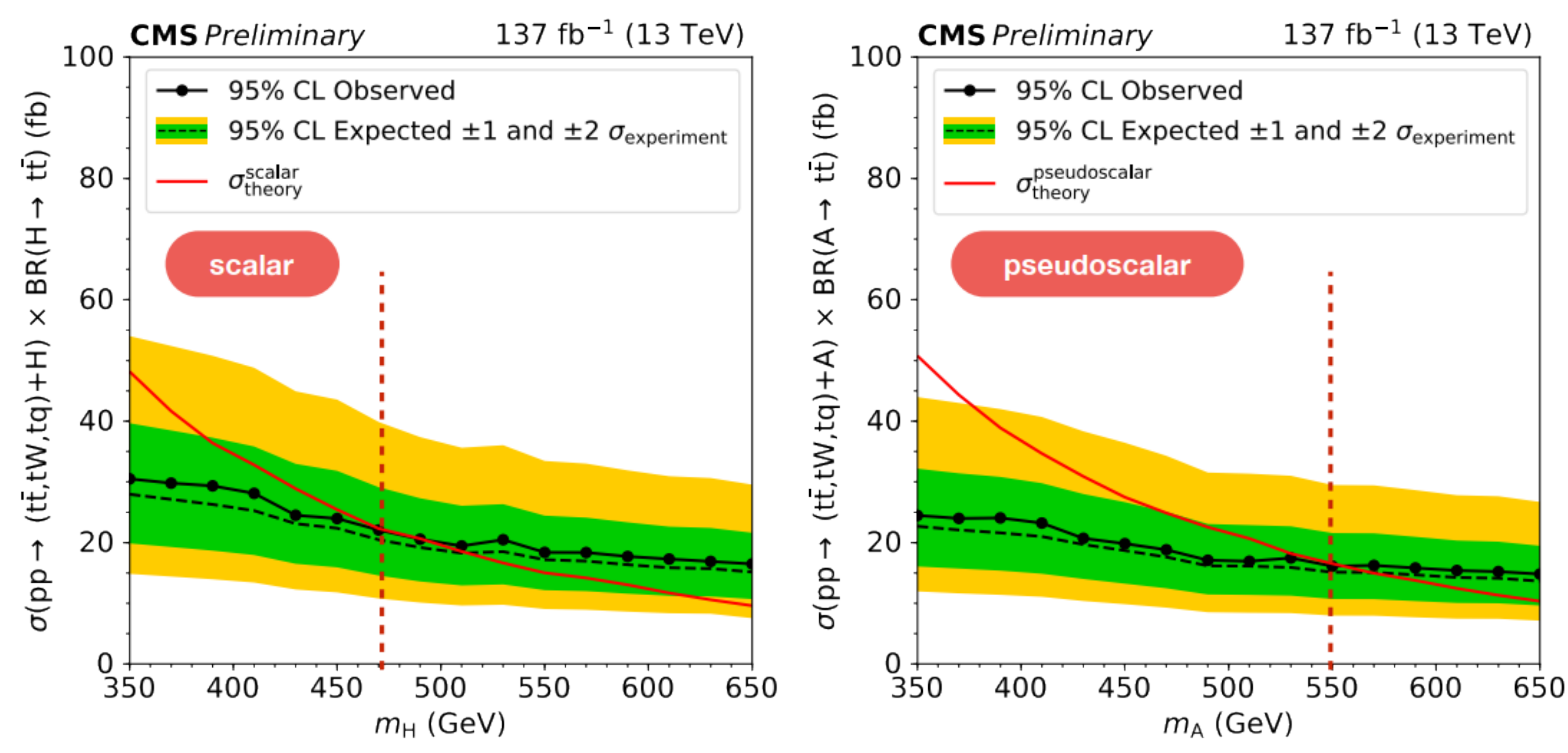}
\caption{The observed and expected 95\% CL upper limits on cross section times branching ratio to $t\bar{t}$ of the production of new scalar (left) and pseudoscalar (right) heavy Higgs as a function of mass. The theoretical value for the 3HDM in the alignment limit is plotted on the solid line.\cite{1908.06463}}
\label{fig:2hdm}
\end{figure}

New particles with $m>2m_t$ that couple to the top quark can also be constrained by a measurement of $\sigma\left(t\bar{t}t\bar{t}\right)$. In particular, we considered the Type-II Two Higgs Doublet Model (2HDM)~\cite{Dicus:1994bm, Craig:2015jba, Craig:2016ygr}. A general 2HDM predicts four new ``Higgs'' particles, but in the ``alignment condition'' the lightest CP-even particle becomes the Standard Model Higgs. Of the remaining new particles, we consider a heavy scalar and heavy pseudoscalar which couple to top quarks similarly to the SM higgs. Figure~\ref{fig:2hdm} shows the cross section times branching ratio of the scalar (H) and pseudoscalar (A) particles as a function of their respective masses as well as the observed limit for our analysis. We exclude a new scalar (pseudoscalar) with mass below 470 (550) GeV at the 95\% confidence level.

\section{Conclusion}

The measurement of the $t\bar{t}t\bar{t}$ cross section is an important tool in better understanding interesting aspects of the Standard Model, as well as an important source of constraint on several BSM theories. Additional interpretations have been considered and are detailed in the paper\cite{1908.06463}. This cross section measurement is found to be consistent with the Standard Model prediction. However, the additional data and increased center-of-mass energy promised by the HL-LHC makes for an exiting future of precision measurements of $t\bar{t}t\bar{t}$ and other rare Standard Model processes\cite{azzi2019standard}.

\section*{Acknowledgements}

I would like to recognize my co-analyzers: Nick Amin, Claudio Campagnari, Frank Golf, and Giovanni Zevi Della Porta. Their ingenuity and willingness to help this beginner was and continues to be greatly appreciated.

\end{document}